\documentstyle[fleqn]{article}

\begin{document}

\title{The off-shell electromagnetic vertex of the nucleon
in chiral perturbation theory}

\author{J.W. Bos\thanks{email: jwb@paramount.nikhefk.nikhef.nl}\\
National Institute for Nuclear Physics and High Energy \\
Physics Section K (NIKHEF-K) \\
P.O. Box 41882, 1009 DB Amsterdam, The Netherlands \\
\\
and
\\
\\
J.H. Koch\\
NIKHEF-K \\
and \\
Institute for Theoretical Physics\\
University of Amsterdam\\
1018 XE Amsterdam, The Netherlands \\
\\
\\
NIKHEF 93-P1, To be published in Nucl. Phys. {\bf A} (1993)}

\date{}

\maketitle

\begin{abstract}

We study the electromagnetic vertex of a  nucleon
in next-to-leading order chiral perturbation theory (CPT).
We consider the case where   one of the
nucleons at the $\gamma$NN vertex is off its mass shell.
We define relevant measures for the off-shell
dependence in the limited kinematical range allowed,
and analyze their expansion in the pion mass.
The leading nonanalytic contributions are calculated to estimate
their size.

\end{abstract}

\section{Introduction}

Many investigations have been carried out recently to examine the
electromagnetic
interaction of a nucleon not on its mass shell. This was mainly done by
using nuclear reactions such as inclusive electron scattering, (e,e$'$), and
nucleon knock-out, (e,e$'$p). However, it also plays
a role, under much simpler circumstances, in two-step reactions
on a free nucleon, such as
pion photo- and electroproduction and Compton scattering.
The electromagnetic vertex in these reactions involves a
nucleon which does not satisfy the free Dirac equation, and consequently,
has a four momentum with $p^2 \ne M^2$.
Therefore, it is not correct to use the {\em free} nucleon electromagnetic
current operator. The operator structure of
the off-shell $\gamma$NN vertex is more
complicated: Its most general form
consists of twelve Dirac operators, each multiplied by a
form factor which depends on three scalar variables \cite{binc60}.
Clearly, these off-shell form factors cannot directly be obtained
from experimental data.
There have been some efforts to obtain them using
dispersion relations \cite{binc60,nyma70,hare72}.
However, because of bad convergence and the
absence of the necessary experimental
input, this approach was of little use in practice. A microscopic model
\cite{naus87,tiem90,song92} for
the internal nucleon structure is needed to describe the full vertex.

In this paper we examine the general electromagnetic vertex of the
nucleon in the framework of chiral perturbation theory (CPT);
since we are interested in off-shell effects we must go at least
to next-to-leading order, which includes one-loop contributions.
A detailed outline of CPT with pions and nucleons, in the relativistic
framework, can be found in Refs.~\cite{gass88,krau90,meis92}.
One starts with the  most general gauge
invariant lagrangian in terms of baryon and meson degrees
of freedom, employing the chiral symmetry of
the underlying QCD lagrangian in the massless limit.
Based on this lagrangian, a perturbation scheme is then developed in which
one expands in the (three) momenta of the external particles and
simultaneously in the mass of the Goldstone bosons.
It has already been applied to pion electro- and
photoproduction \cite{bern91,bern92.1} and Compton scattering at
threshold \cite{bern92.3,butl92}. Certain aspects of the off-shell vertex
we study here are already implicitly contained in these low-energy results. We
use the relativistic framework; for the heavy baryon limit of CPT we
refer e.g. to Ref.~\cite{jenk92}.

\section{General aspects of the off-shell  $\gamma$NN vertex}

The general form of the $\gamma$NN vertex
for the case that both
initial and final nucleon are off their mass shell,
has been given by Bincer \cite{binc60}. Using the
projection operators
\begin{equation}
	\Lambda^{\pm}(p)=\frac{\pm p\!\!\hspace{-0.45mm}\slash + W}{2W},
\end{equation}
with $W=+\sqrt{p^2}$,
this vertex can be rearranged as
\begin{equation}
	\Gamma^{\rm{N}}_{\mu}(p',p) =
	-{\rm i} e \sum_{\alpha ,\beta =-}^{+}
	\Lambda^{\alpha}(p') \Bigl[ f_{1}^{\alpha \beta}
	\gamma_{\mu} + f_{2}^{\alpha \beta} \frac{{\rm i} \sigma_{\mu \nu}
	q^{\nu}}{2M}
	+ f_{3}^{\alpha \beta} q_{\mu} \Bigr]
	\Lambda^{\beta}(p).
\label{4:emvert}
\end{equation}
The initial and final nucleon have four-momentum
$p$ and $p'$, respectively, and
the photon four-momentum is $q=p'-p$.
The twelve form factors
$f_{k}^{\alpha \beta}$ each depend on the three scalar variables,
which are usually  chosen as $W=\sqrt{p^2}$, $W'=\sqrt{p'^2}$ and $q^2$.
For simplicity we consider here the situation where the final nucleon
is on-shell, $p'^2 = M^2$, while the initial nucleon (before
absorption of the photon) is off its mass shell.
This ``half off-shell'' vertex can be easily obtained from
eq.~(\ref{4:emvert}):
\begin{eqnarray}
	\overline{\rm u}(p') \Gamma^{\rm{N}}_{\mu}(p',p) & = &
	- {\rm i} e \overline{ \rm u}(p') \Bigl[
	( f_{1}^{++}
	\gamma_{\mu} + f_{2}^{++} \frac{{\rm i} \sigma_{\mu \nu}
	q^{\nu}}{2M}
	+ f_{3}^{++} q_{\mu} )
	\Lambda^{+}(p)
\nonumber
\\
	& & \mbox{} +
	( f_{1}^{+-}
	\gamma_{\mu} + f_{2}^{+-} \frac{{\rm i} \sigma_{\mu \nu}
	q^{\nu}}{2M}
	+ f_{3}^{+-} q_{\mu} )
	\Lambda^{-}(p) \Bigr],
\label{4:hovert}
\end{eqnarray}
where $f_{i}^{+\pm} = f_{i}^{+\pm} (q^2,W,M)$.
We will from now on work with the {\em reducible\/} vertex operator, which
includes self-energy insertions into the external nucleon legs.
Taking the final nucleon on-shell, we need to insert the self-energy
into the initial nucleon leg only.

In order to show
the connection between the general form factors in
eqs.~(\ref{4:emvert}) and~(\ref{4:hovert}) and the usual free form factors, we
evaluate $\Gamma^{\rm{N}}_{\mu}(p',p)$
between free on-shell nucleon spinors.
Using eq.~(\ref{4:hovert})  one finds for the Dirac form factor \cite{binc60}
\begin{equation}
	F_1 (q^2)  =  f_1^{++} (q^2,M,M),
\label{4:dirac}
\end{equation}
and for the  Pauli form factor
\begin{equation}
	F_2 (q^2)  =  f_2^{++} (q^2,M,M).
\label{4:pauli}
\end{equation}
Even though they do not appear in the free current, we will
also use the analogously defined functions
\begin{equation}
	{\cal F}_1 (q^2)  =  f_1^{+-} (q^2,M,M),
\label{4:merel1}
\end{equation}
and
\begin{equation}
	{\cal F}_2 (q^2)  =  f_2^{+-} (q^2,M,M).
\label{4:merel2}
\end{equation}
Below, we will only discuss the four
half off-shell form factors $f_1^{+\pm}$ and $f_2^{+\pm}$, since
$f_3^{+\pm}$ is directly related to $f_1^{+\pm}$ through the
Ward-Takahashi identity \cite{binc60}.

\section{The $\gamma$NN vertex in CPT}

We will calculate the electromagnetic interaction term in next-to-leading
order CPT, i.e.\ up to order ${\cal O}( {E^3})$, where $E$ is a
generic notation for the four-momentum of the external photon, the
three-momenta of the external nucleons, and the meson mass \cite{gass88}.
For this purpose we need the CPT lagrangian up to order ${\cal O}({E^3} )$,
which reads
\begin{equation}
	{\cal L} = {\cal L}_1 + {\cal L}_2,
	\label{4:CPT}
\end{equation}
where the first part, ${\cal L}_1$, is the standard
nonlinear $\sigma$-model lagrangian.
We restrict ourselves
here to the SU(2) flavor group and
the isospin symmetric limit where $m_{\rm{u}} = m_{\rm{d}} = m$,
explicitly only considering pions and nucleons.
In Refs.~\cite{gass88}~and~\cite{krau90} the second part
of the lagrangian eq.~(\ref{4:CPT}), ${\cal L}_2$, is given sufficient
for the situation where one considers on-shell matrix elements
of the $\gamma$NN vertex. For
off-shell particles, however, one needs to take into account
more terms.
For our situation the necessary expression is, in the notation
of  Ref~\cite{gass88},
\begin{equation}
	{\cal L}_2  = {\cal L}_2^{(0)}  + {\cal L}_2^{(2)} + {\cal L}_2^{(3)},
	\label{4:lag2}
\end{equation}
where
\begin{eqnarray}
	{\cal L}_2^{(0)} & = & \Delta M \frac{M^2}{F^2} \bar{\psi} \psi
\\
	{\cal L}_2^{(2)} & = & c_1 \frac{4M_{\pi}^2 M}{F^2} \bar{\psi} \psi +
	c_6 \frac{M}{F^2} \bar{\psi} \sigma^{\mu \nu} f_{\mu\nu}^+ \psi +
	n_1 \frac{M}{F^2} \bar{\psi} ({\rm i} D\!\!\!\!\slash - M)^2 \psi
\label{4:kieviet}
\\
	{\cal L}_2^{(3)} & = &  b_9 \frac{1}{F^2} \bar{\psi} \gamma^{\mu} [ D^{\nu},
	f_{\mu\nu}^+ ] \psi + b_{12} \frac{4M_{\pi}^2}{F^2} \bar{\psi} ({\rm i}
	D\!\!\!\!\slash - M ) \psi
\nonumber
\\
	 & & \mbox{} +
	n_2 \frac{1}{F^2} \bar{\psi} \{ ({\rm i} D\!\!\!\!\slash - M),
	\sigma^{\mu \nu} f_{\mu\nu}^+ \} \psi
	 + n_3 \frac{1}{F^2} \bar{\psi} ({\rm i} D\!\!\!\!\slash - M)^3 \psi.
\label{4:mus}
\end{eqnarray}
Here $D_{\mu}$ is the covariant derivative and $f_{\mu\nu}^+$ is
the electromagnetic field strength tensor.
The ``low-energy constants'' $\Delta M$, $c_1$, \ldots, $n_3$ in
${\cal L}_2$ are  parameters
needed for the renormalization of the
ultra-violet divergences.
The constants $n_1$, $n_2$, and $n_3$ in eqs.~(\ref{4:kieviet})
and~(\ref{4:mus})
only appear in the half off-shell electromagnetic vertex,
which is not a direct observable. In a  {\em physical\/}  amplitude
where this half off-shell vertex plays a role, e.g.\
Compton scattering on a free nucleon, such terms are
included through counterterms for the entire amplitude.

For the evaluation of the $\gamma$NN vertex in next-to-leading order we
need diagrams from ${\cal L}_1$ up to
the one-loop level (shown in fig.~1) and from ${\cal L}_2$ only on the
tree-level.
\begin{figure}
\caption{\em Loop diagrams contributing to the half off-shell electromagnetic
vertex.}
\end{figure}
Applying the standard Feynman rules, and using
the renormalization prescription of dimensional
regularization, one obtains for example
for diagram~(1d):
\begin{eqnarray}
	\Gamma^{\rm{d}} _{\mu} & = &
	M_{\rm ref}^{4-n}\!\!\int\!\!\frac{d^n k}{(2\pi)^n}
	\frac{g_{\rm \scriptscriptstyle A}}{2F}
	( - k \!\!\!\slash - q\!\!\!\slash  ) \gamma_5
	 \tau_i  \Delta(k+q)
\nonumber
\\*
	& &  \times \Bigl[ -e \epsilon^{ij3}
	(2k+q) _{\mu}  \Bigr]
	 \Delta(k) S(p-k) \frac{g_{\rm \scriptscriptstyle A}}{2F}
	k \!\!\!\slash \gamma_5
	 \tau_j.
\label{4:diagram}
\end{eqnarray}
Here $S$ and $\Delta$ are the bare nucleon and
pion propagator, respectively.
As this example shows, the loop diagrams diverge for $n=4$.
Finite results are obtained by using the counterterms
from ${\cal L}_2$, eq.~( \ref{4:lag2}).
The constants $n_1$ and $n_3$ are needed to renormalize
the off-shell {\em irreducible\/} $\gamma$NN vertex and the nucleon
self-energy.
However, in the {\em reducible\/} vertex we consider here they drop out.
We will later discuss the relevance of the constant $n_2$ for
our situation; the other counterterms,
which enter when renormalizing the on-shell (observable)
properties,  were already discussed in Ref.~\cite{gass88}.

Let us now discuss to what extent we can obtain the half off-shell
electromagnetic form factors in eq.~( \ref{4:hovert}) with this next-to-leading
order calculation.
Since $A_{\mu} = {\cal O}( {E})$, the
$\gamma$NN vertex can be evaluated
up to ${\cal O}( E^2)$, yielding a total interaction term of
${\cal O}({E^3} )$.
In this expansion scheme the operators
in  eq.~(\ref{4:hovert}) are counted as
\begin{equation}
	\Lambda^+ (p) = {\cal O}( {1}); \, \Lambda^- (p) =
	{\cal O}( {E}); \,  q^{\nu} = {\cal O}(E).
\end{equation}
It implies that we can obtain $f_1^{++}$ up to terms
of ${\cal O}( {E^2})$,
$f_1^{+-}$ and $f_2^{++}$ up to terms of ${\cal O}( {E})$,
and $f_2^{+-}$  up to a constant.
Since for the scalar variables one has
\begin{equation}
	W-M = {\cal O}( E); \, q^2 = {\cal O}({E^2}),
\end{equation}
we may only use the calculation in a limited kinematical region, where
$W$ is close to $M$ and $q^2$ is small.
As we will see, however, the expansion in $E$
also contains   nonanalytic terms in the quark mass
(or in the pion mass, $M_{\pi}$, since
$ M_{\pi}^2 \sim m$).
It is generally
assumed that they can uniquely be obtained
from a one-loop CPT calculation \cite{gass88,meis92}.
In this paper, we will focus
on the leading singular terms  in the chiral limit,
$M_{\pi} \rightarrow 0$. Since the actual pion mass is small
these contributions can be expected to be the dominant.

Expanding around the photon point, one has for a given form factor
\begin{equation}
	f(q^2,W,M) = f(0,W,M) +
	\frac{\partial}{\partial q^2}f(q^2,W,M)|_{q^2 = 0} + \ldots\, .
\label{4:paard}
\end{equation}
For $f_1^{++}$ and $f_1^{+-}$, the first term in such an expansion is defined
by the nucleon charge, independent of $W$ \cite{binc60}.
The $W$ dependence will first show up in  the second
term, and we will therefore below consider
\begin{equation}
	6 \frac{\partial^2}{\partial W\,\partial q^2}
	f_1^{+\pm} (q^2,W,M) |_{W=M,q^2=0}.
\label{4:q1}
\end{equation}
We have included the factor 6
since for $f_1^{++}$ the second term in the right hand side of
eq.~(\ref{4:paard}) is related
to the mean square radius of the nucleon according to
\begin{equation}
	\langle r^2 \rangle =  6 \frac{d}{dq^2}
	F_1 (q^2)|_{q^2=0}.
\label{4:giraf}
\end{equation}
Therefore, the quantity in eq.~(\ref{4:q1}) for $f_1^{++}$
can be seen as the change of the mean square radius
with $W$.

Since for $f_2^{++}$ and $f_2^{+-}$
already the first term of the expansion eq.~(\ref{4:paard}) is
$W$ dependent \cite{binc60,naus87}, we will consider the
leading off-shell variation
\begin{equation}
	 \frac{\partial }{\partial W} f_2^{+\pm} (0,W,M) |_{W=M}.
\label{4:q2}
\end{equation}

If there are nonanalytic terms in the quantities defined
in eqs.~(\ref{4:q1}) and~(\ref{4:q2}) then they are determined by a
one-loop calculation. According to the above power counting
rules, the analytic terms for these properties for
$f_1^{++}$, $f_1^{+-}$ and $f_2^{+-}$
fall outside the scope of this paper.

\subsection{The four form factors at $W=M$}

Before discussing the above off-shell properties,
we first will examine the four form factors defined
in eqs.~(\ref{4:dirac} ) to~( \ref{4:merel2}), i.e.\ $F_1$, ${\cal F}_1$,
$F_2$,  and
${\cal F}_2$.

\begin{enumerate}

\item
The relevant feature for the charge form factor
$F_1$, for low $q^2$,
is the mean square radius defined in eq.~(\ref{4:giraf}).
Its {\em isovector\/} part has already been calculated by Gasser et al
\cite{gass88}. They found
\begin{equation}
	\langle r^2 \rangle^{\rm V} = a_{-1} \ln x + a_0 + {\cal O}({\sqrt{x}} ),
\label{4:aap1}
\end{equation}
where $x = M_{\pi}^2 / M^2$ is linear in the quark mass.
CPT fixes the coefficient of the leading nonanalytic contribution,
\begin{equation}
	a_{-1} = - \frac{1}{16 \pi^2}\frac{1}{ F^2} (1 + 5 g_{\rm \scriptscriptstyle
A}
	^2)
\label{4:kameel}.
\end{equation}
However, the next term in  eq.~(\ref{4:aap1}), $a_0$, contains
the low-energy parameter $b_9^{\rm V}$, appearing in the
isovector part of eq.~(\ref{4:mus}).
Numerically, the leading term in eq.~(\ref{4:aap1}) gives an
isovector mean radius
of $1.0 \; {\rm fm}$, while the
experimental value is $0.8 \; {\rm fm}$.
The constant $a_0$  can be
fitted to exactly arrive at the experimental value
(which amounts to fixing $b_9^{\rm V}$).

We find for the {\em isoscalar\/} part of the mean square radius
\begin{equation}
	\langle r^2 \rangle^{\rm S} = b_0 + {\cal O}( {\sqrt{x}}),
\end{equation}
where $b_0$ is determined by the
isoscalar constant $b_9^{\rm S}$ in eq.~(\ref{4:mus}). Therefore,
in contrast to $\langle r^2 \rangle^{\rm V}$,
the leading part of $\langle r^2 \rangle^{\rm S}$ is not a
singular term. The constant $b_0$ can be
used to fit the experimental value, which fixes $b_9^{\rm S}$.

\item
The slope of the {\em isovector\/} form factor ${\cal F}_1^{\rm V}$ at the
photon point is given by
\begin{equation}
	6 \frac{d}{dq^2} {\cal F}_1^{\rm V}(q^2)|_{q^2 = 0} =
 	c_{-1} \ln x + {\cal O}( {1}),
\label{4:msmp}
\end{equation}
where the coefficient of the leading singular term is
\begin{equation}
	c_{-1} = - \frac{1}{16 \pi^2}\frac{1}{ F^2}
	 (1 - g_{\rm \scriptscriptstyle A}^2).
\end{equation}
Since $g_{\rm \scriptscriptstyle A} = 1.25$, this is a much smaller leading
coefficient than $a_{-1}$
in eq.~(\ref{4:aap1}), indicating that
${\cal F}_1^{\rm V}(q^2)$ has a weaker $q^2$ dependence than
$F_1^{V}$ in the chiral limit.
Consequently, the higher order terms in eq.~(\ref{4:msmp})
can be important, but
are outside the scope of our ${\cal O}( {E^2})$ calculation of
the vertex.

The slope of the {\em isoscalar\/} form factor
${\cal F}_1^{
\rm S} (q^2)$ has no singular leading contributions.
{}From the operator structure one can see that an analytic
contribution (starting with a constant) will lead
at least to an ${\cal O}( {E^3})$ contribution to the vertex. Again, this
is outside the scope of our calculation.

\item
More terms in the expansion in $x$ of the {\em isovector\/} anomalous
magnetic moment, $F^{\rm V}_2 (0)$, can be obtained with
our calculation.
The expansion is given by \cite{gass88}
\begin{equation}
	F_2^{\rm V} (0)  = 8 c_6^{\rm{V}} \frac{M^2}{F^2} +
	\frac{g_{\rm{A}}^2}{16 \pi^2}
	\frac{M^2}{F^2}
	\Bigl[ 5 - 4 \pi \sqrt{x} - 7 x \ln x  + {\cal O}( {x}) \Bigr],
\label{4:koe1}
\end{equation}
while we find for the {\em isoscalar} part
\begin{equation}
	F_2^{\rm S} (0) = 8 c_6^{\rm{S}} \frac{M^2}{F^2}
	- \frac{g_{\rm{A}}^2}{16 \pi^2} \frac{M^2}{F^2}
	\Bigl[ 3 + 3 x \ln x  + {\cal O}( {x}) \Bigr].
\label{4:koe2}
\end{equation}
Thus, both the isoscalar and isovector
anomalous magnetic moment remain finite in
the chiral limit.
Their leading contribution, the constant term
in eqs.~(\ref{4:koe1}) and~(\ref{4:koe2}),
already contain the free constants $c_6^{
\rm S}$ and $c_6^{\rm V}$, which
one can use to fit the experimental values.

\item
At the photon point the $+-$ form factors ${\cal F}_2^{\rm S}$ and
${\cal F}_2^{\rm V}$ are simply
given by
\begin{equation}
	{\cal F}_2^{\rm S/V} (0) = 8 c_6^{\rm S/V} - 8 n_2^{\rm S/V} +
	{\cal O}( {x}),
\label{4:ampm}
\end{equation}
i.e. their leading contributions
are determined by the low-energy constants from ${\cal L}_2$.
In contrast to the low-energy constants $c_6^{\rm S}$
and $c_6^{\rm V}$, which enter
the anomalous magnetic moments eqs.~(\ref{4:koe1}) and~(\ref{4:koe2})
and can be determined by fitting this property,
$n_2^{\rm S}$ and $n_2^{\rm V}$ in eq.~(\ref{4:ampm}) do not enter
the on-shell nucleon matrix elements.
Stated differently, to obtain the leading
contributions to ${\cal F}_2^{\rm S/V}(0)$ we need more
information than provided by
chiral symmetry and on-shell nucleon properties alone.

\end{enumerate}

Note that an unsubtracted dispersion relation would lead to \cite{nyma70}
\begin{equation}
	F_2^{\rm S/V} (0) \approx
	\frac{1}{2} \sqrt{x} {\cal F}_2^{\rm S/V} (0) ,
\end{equation}
which is not satisfied as can be seen
by comparing the powers of $\sqrt{x}$ in
eq.~(\ref{4:ampm}) and in eqs.~(\ref{4:koe1}) and~(\ref{4:koe2}).
Nyman had already questioned the validity of this approach
in his study of p-p brems\-strahlung \cite{nyma71};
the above provides another reason.
such an approach would fail.

Therefore, for the two quantities ${\cal F}_1$ and ${\cal F}_2$, which
do not occur in the free on-shell current, we cannot extract
any definitive information. While the leading singular
term for ${\cal F}_1^{\rm V}$ is determined, it is unlikely that it will
be numerically significant.

\subsection{Off-shell dependence of the form factors}

We will now turn to the off-shell
behavior of the form factors, in particular the quantities
defined by eqs.~(\ref{4:q1}) and~(\ref{4:q2}).

\begin{enumerate}

\item
For the {\em isovector\/} form factors $f_1^{++, {\rm V}}$
and $f_1^{+-, {\rm V}}$ we find at $q^2 = 0$ and $W=M$
\begin{equation}
	6 \frac{\partial^2}{\partial W \partial q^2} f_1^{++, {\rm V}}  =
	\frac{5}{32 \pi}
	\frac{g_{\rm \scriptscriptstyle A}^2}{F^2 M} \frac{1}{\sqrt{x}}
	+ {\cal O}( {\ln x}),
\label{4:geit1}
\end{equation}
and
\begin{equation}
	6 \frac{\partial^2}{\partial W \partial q^2} f_1^{+-, {\rm V}}  =
	-\frac{1}{16 \pi}\frac{g_{\rm \scriptscriptstyle A}^2}{F^2 M}
	\frac{1}{\sqrt{x}} + {\cal O}({\ln x} ).
\label{4:geit2}
\end{equation}
Therefore, in the chiral limit both quantities diverge.
{}From eq.~(\ref{4:geit1}) we can see
that the isovector mean square radius of a nucleon
with $W<M$ is smaller
than that of a free one.
This is in qualitative agreement with the results from
Refs~\cite{naus87} and~\cite{tiem90}.
Taking $x$ at its physical value, $x = 0.022$, and retaining only
the leading terms in eq.~(\ref{4:geit1}), we find in CPT that the free
($W = 940 \; {\rm MeV}$)
mean square radius of $0.6 \; {\rm fm}^2$ is reduced
by about $0.1 \; {\rm fm}^2$ if we use eq.~(\ref{4:geit1}) to extrapolate
linearly to $W = 890 \; {\rm MeV}$.
The above illustrates that the ``radius of an off-shell
nucleon'' has not a fixed value, but
is a dynamic property  depending on the kinematical
variables (in our case, where the nucleon is half off-shell, this
variable is $W$).
Comparing eqs.~(\ref{4:geit1}) and~(\ref{4:geit2})
indicates that the $W$ dependence
of $f_1^{+-, \rm V}$ is smaller than that of $f_1^{++, {\rm V}}$.
It is interesting that the slope of $f_1^{+-, {\rm V}}$ near
the photon point increases
for $W < M$, in contrast to that of $f_1^{++, {\rm V}}$.

\item
The singular terms for the {\em isoscalar\/} form factors
$f_1^{++, {\rm S}}$ and $f_1^{+-, {\rm S}}$ vanish:
\begin{equation}
   6 \frac{\partial^2}{\partial W \partial q^2} f_1^{+\pm, {\rm S}} = {\cal
O}({1} ).
\end{equation}
Their expansions start with a ${\cal O}( {1})$ term, indicating a weaker
off-shell dependence than the  isovector parts.
As already discussed, these analytic contributions (which are now
the leading one) cannot
be obtained from our calculation.

\item
The off-shell variation of the
{\em isovector\/} form factors $f_2^{++, {\rm V}}$
and $f_2^{+-, {\rm V}}$ at $q^2 = 0$ and $W=M$ reads
\begin{equation}
	\frac{\partial}{\partial W} f_2^{++, {\rm V}}
	 = d_{-1} \ln x + d_0 + {\cal O}( {\sqrt{x}})
\label{4:schaap1}
\end{equation}
and
\begin{equation}
	\frac{\partial}{\partial W} f_2^{+-, {\rm V}}
	 =  - d_{-1} \ln x + {\cal O}( {1}),
\label{4:schaap2}
\end{equation}
respectively, where
\begin{equation}
	d_{-1} = - \frac{g_{\rm \scriptscriptstyle A}^2}{8 \pi^2} \frac{M}{F^2},
\end{equation}
and $d_0$ is a constant containing $n_2^{\rm V}$.
Note that, according to eq.~(\ref{4:ampm}), $n_2^{\rm V}$
also determines the on-shell
limit of $f_2^{+-,{\rm V}}$.
In contrast to their values at the photon point, given in
eqs.~(\ref{4:koe1}) and~(\ref{4:ampm}),
the dependence of $f_2^{++, \rm V}$ and $f_2^{+-, \rm V}$
on $W$ are fixed, in leading order.
The leading singular terms in eqs.~(\ref{4:schaap1}) and~(\ref{4:schaap2})
are of equal magnitude but
have opposite sign.
Retaining only the leading contribution in eq.~(\ref{4:schaap1}),
we find that if we extrapolate  to $W = 890 \; {\rm MeV}$ the off-shell
isovector anomalous magnetic moment
decreases by $0.4$, about $10$ \% of the on-shell value.

\item
Again, the derivatives of the {\em isoscalar\/} form factors
$f_2^{++,{\rm S}}$ and $f_2^{+-,{\rm S}}$ with respect to $W$
are not singular in the chiral
limit.
Since the vertex operators multiplying $f_2^{+-,{\rm S}}$ in the
vertex are already of order ${\cal O}( {E^2})$ the absence of a
singular term means that we cannot make a prediction
for this term. The operators for $f_2^{++,{\rm S}}$ are of
order ${\cal O}( {E})$.
The expansion of the derivative starts off with a constant term
containing $n_2^{\rm S}$, not
 further restricted by chiral
symmetry. As already discussed in connection with ${\cal F}_2^{
\rm S} (0)$,
eq.~(\ref{4:ampm}),
this constant is not constrained by on-shell properties
of the nucleon. Therefore, no prediction result for the
off-shell variation of both isoscalar form factors.

\end{enumerate}
In the above, we have only been able to determine off-shell
effects through nonanalytic terms in the isovector part
of the vertex. These effects are entirely due to the
contribution from the loop diagram (d) in fig.~1, given
by eq.~(\ref{4:diagram}), which is purely isovector.

\section{Summary}

A way to address the electromagnetic interaction of an off-shell nucleon
in a dynamical fashion is chiral perturbation theory.
It is a general approach based on  the fundamental QCD symmetries.
We are interested in the off-shell behavior of the reducible
electromagnetic vertex.
This means in CPT that one has to expand the lagrangian
at least to ${\cal O}( {E^3})$, where
$E$ is a generic nucleon three-momentum, photon four-momentum or a
meson mass. We restrict ourselves to an ${\cal O}( {E^3})$ calculation.
First, this means that one has to take into account one-loop
contributions due to iterating the lowest order term in the lagrangian.
It is generally assumed that the one-loop diagrams yield
all the non-analytic terms in the expansion due to
the expansion in the pion mass. Since the pion mass is small, one
can expect that these  nonanalytic contributions provide
a relevant part of the full expansion. Second, the higher order terms
(up to ${\cal O}({E^3} )$) in
the lagrangian also must be taken into account,
but  only at the tree level. There are also
contributions in this order from higher loops; they are
effectively contained in the  parameters of the
lagrangian, which are not predicted by the theory.
The parameters that appear in the on-shell case can be used to
fit the free nucleon properties.

In   an expansion to this order
the range of kinematical variables of
the vertex is restricted.
In practice, this means that e.g.\ only the slope of the
form factors can be studied.
We have, for simplicity, studied the
``half off-shell'' electromagnetic
vertex, i.e.\ the situation where only one of the nucleons at
the vertex
is off its mass shell. The corresponding
reducible form factors $f_1^{++}$ and $f_1^{+-}$
are restricted to the nucleon charge at the photon point, independent of $W$.
Therefore, off-shell features in $f_1^{++}$ and $f_1^{+-}$ first
show up in the slope of these form factors, e.g.\ for $f_1^{++}$
as the $W$ dependence
of the mean square radius. Our calculation shows that in general
only the isovector off-shell form factors contain singular
contributions which can be determined already at the one-loop level. For
$f_1^{++,{\rm V}}$ this contribution indicates a reduction
of the free isovector mean square radius when $W<M$; for example
a decrease of about 15 \% at $W=890 \; {\rm MeV}$. For $f_1^{+-,{\rm V}}$,
it increases for $W<M$, but this effect is much
smaller.

For the half off-shell form factors $f_2^{++}$ and $f_2^{+-}$ a
dependence on the invariant mass $W$ already occurs at the
photon point. Decreasing $W$ from the free mass to $890 \; {\rm MeV}$
we obtain from the leading terms a decrease of $f_2^{++, {\rm V}}$
at the photon point, while $f_2^{+-, {\rm V}}$ increases by
that amount. For the half off-shell isoscalar derivatives one does not
have a leading singular contribution. The analytic terms depend
on low-energy constants which only enter in the electromagnetic
vertex involving an off-shell nucleon, and therefore cannot be
determined from on-shell properties.

In our approach, we have worked in the nonstrange sector.
Furthermore, only the pion, the lightest meson
appears, i.e. only the ``soft'' degrees of freedom are explicitly
taken into account. The ``hard'' physics due to heavier
mesons appears through the choice of the renormalization constants
as was argued, e.g.\ in connection with the renormalization constant
$c_6$ \cite{meis92}.
In this paper, we have focussed on the leading singular terms
in the half off-shell form factors.
The exact value of the quantities considered
depends on the size of the higher-order contributions. However,
our calculations were able to show  qualitative
features of the electromagnetic properties of a nucleon not on
its mass shell. Our results once again show that these properties,
such as the ``radius of an off-shell nucleon,'' are not fixed
properties, but depend on the kinematical parameters and dynamical
circumstances.

\section*{Acknowledgement}

We would like to thank S. Scherer for useful criticism.
This work was made possible by financial support from the Foundation
for Fundamental Research on Matter (FOM) and the Netherlands
Organization for
Scientific Research (NWO).

\end{document}